\documentclass[pra,onecolumn,floatfix,a4paper,superscriptaddress]{revtex4}
\usepackage{bm,color,graphicx,amsmath,txfonts}

\usepackage[colorlinks, citecolor=blue,linkcolor=blue]{hyperref}

\newcommand{\diag}{{diag}}

\begin{document}

\title{Enhancing One-Way Steering and Non-Classical Correlations in Magnomechanics via Coherent Feedback}

\author{Hamza Harraf}
\affiliation{LPHE-Modeling and Simulation, Faculty of Sciences, Mohammed V University in Rabat, Rabat, Morocco.}	
\author{Noura Chabar}
\affiliation{LPTHE-Department of Physics, Faculty of Sciences, Ibnou Zohr University, Agadir 80000, Morocco}
\author{Mohamed Amazioug} \thanks{amazioug@gmail.com}
\affiliation{LPTHE-Department of Physics, Faculty of Sciences, Ibnou Zohr University, Agadir 80000, Morocco}
\author{Rachid Ahl Laamara}
\affiliation{LPHE-Modeling and Simulation, Faculty of Sciences, Mohammed V University in Rabat, Rabat, Morocco.}
\affiliation{Centre of Physics and Mathematics, CPM, Faculty of Sciences, Mohammed V University in Rabat, Rabat, Morocco.}

\begin{abstract}

In this work, we propose a theoretical scheme to explore the enhancement of quantum correlation hierarchies in a cavity magnonmechanical system via the coherent feedback tool. We use Gaussian geometric discord to quantify quantum correlations between the two magnon modes, including those beyond entanglement, in the steady state. Logarithmic negativity and Gaussian quantum steering are employed to characterize entanglement and steerability, respectively. Our results show that  adjusting the beam splitter's reflective parameter can significantly enhance quantum correlations and increase their resilience to thermal noise. Moreover, we demonstrate that coherent feedback can achieve enhanced genuine tripartite entanglement among the photon, magnon \(M_1\), and phonon. These findings present promising strategies for enhancing entanglement in magnon-based systems and advancing quantum information technologies. We conclude by validating the system and demonstrating its ability to detect entanglement.\\


\end{abstract}

\date{\today}
\maketitle

\section{Introduction}

In recent years, the platform for studying the interaction between the magnon and light has been a subject of enormous interest \cite{zhang2016cavity,li2019entangling,HJ21,asjad23,amazioug2023feedback,S23,nor,amghar}. Magnons, the quanta of collective spin excitations in ferromagnetic crystals  such as  Yttrium Iron  Garnet (YIG) \cite{xiong2023optomechanically}, this quasiparticule can open large possibility for develop the quantum technologies as carriers of quantum information, because it is widely frequency-tunable and can be coherently coupled to different degrees of freedom \cite{osada2016cavity}, the YIG have two special proprieties the first one is the high spin density that participates to interact strongly between microwave cavity photon and magnon creating cavity-magnon polaritons \cite{zhang2015cavity}, and the second one is very low damping rate. The kittel mode \cite{kittel1948theory} in YIG possesses unique proprieties containing interesting magnonic nonlinearities. This strong coupling can have utilizes in quantum information processing because it provides a way to transfer coherent information between radically different information carriers. The field of magnon allows to study numerous interesting topics, like bistability \cite{wang2018bistability}, the attraction of energy levels in cavity magnon-polaritons, progress in cavity spintronics \cite{bai2017cavity}, the appearance of magnon dark modes \cite{wang2018bistability}, by strong coupling of magnon, which has led to the investigation of several elements of quantum information, like the coupling of the superconducting qubit to magnons \cite{tabuchi2015coherent} and phonons \cite{zhang2016cavity}. \\

The entanglement in large scale plays essential role for both information processing and comprehend the quantum theories.  The macroscopic entanglement realizes the boundary between the quantum and classical word \cite{frowis2018macroscopic}, and it possesses many technological applications such as quantum sensing \cite{bennett2000quantum}, quantum networks \cite{matthews2009manipulation}. The entanglement occurs in many systems, and in our situation we concentrate to how that entanglement can be generated in nonlinear quantum systems such as magnomechanical systems \cite{li2018magnon}. In this regard, over the past ten years, cavity magnomechanics has attracted a lot of attention \cite{aspelmeyer2014cavity}, because this system plays essential role for study many phenomena like quantum entanglement  \cite{horodecki2009quantum}. It is hardly to prepare the entanglement between two massive objects because the environment  noise and difficulty to control the large scale systems. Furthermore, decreasing the superposition and entanglement of big objects may possibly be a function of gravitationally caused decoherence \cite{frowis2018macroscopic}. These macroscopic entangled states are claimed to aid in the comprehension of basic questions in modern physics, such as quantum gravity \cite{marletto2017gravitationally}, transition between quantum to classic \cite{frowis2018macroscopic}.\\

In this work, we propose a scheme to prepare two magnon modes of two massive YIG spheres in an entangled state and to enhance bipartite and tripartite quantum correlations in a magnomechanical system. This is achieved by introducing a coherent feedback technique, which improves upon the results obtained in \cite{li2019entangling}. The coherent feedback has the advantages in many aspects: e.g., in cooling \cite{hamerly2015quantum}, entanglement \cite{asjadfb,huangfb,amaziougPLA2020,amazioug2023feedback}, suppressing noises \cite{yanagisawa2003transfer} and quantum network \cite{hein2015entanglement}. The two magnon modes couple to a single microwave cavity mode via linear beamsplitter interactions. Nevertheless, by activating the magnetostrictive (radiation pressure-like) interaction in one YIG sphere, we show that such an interaction can be utilized to generate entanglement between two magnon modes if one of them is suitably
driven by a microwave field. This system studies both bipartite and tripartite entanglement. Further, the magnon-phonon coupling is the source of the entanglement, which then partially transfers to the cavity-magnon and cavity-phonon subsystems. Additionally, the two magnon modes get entangled by the use of the effective state-swap interaction between the cavity and the other magnon modes. We utilize the logarithmic negativity to quantify the bipartite entanglement, and the residual contangle to examine tripartite entanglement in our system. We shall study the amount of quantum correlations distributed between the different components of the magnomechanical systems by employing Gaussian geometric discord. We use the Gaussian quantum steering to characterize the steerability one-way of the two magnon modes.\\

The structure of this article is as follows: The section \ref{sec2} presents the Hamiltonian and quantum Langevin equations (QLEs) of the optomechanical degrees of freedom in presence of corehent feedback. The section \ref{sec3} focuses on the linearization of quantum Langevin equations and covariance matrix (CM) in steady-state. In the section \ref{sec4}, We employ gaussian quantum steering, logarithmic negativity and Gaussian geometric discord to investigate the hierarchy of quantum correlations between two magnons. The  section \ref{sec6}, we discuss investigate the hierarchy of quantum correlations under the influence of various parameters. Lastly in section \ref{sec7}, we provide an overview of the entire work presented in this article and summarize the results we have obtained.

\section{Model} 
\label{sec2}
\begin{figure}[!h]
\centering
\includegraphics[width=1\linewidth, height=0.37\linewidth]{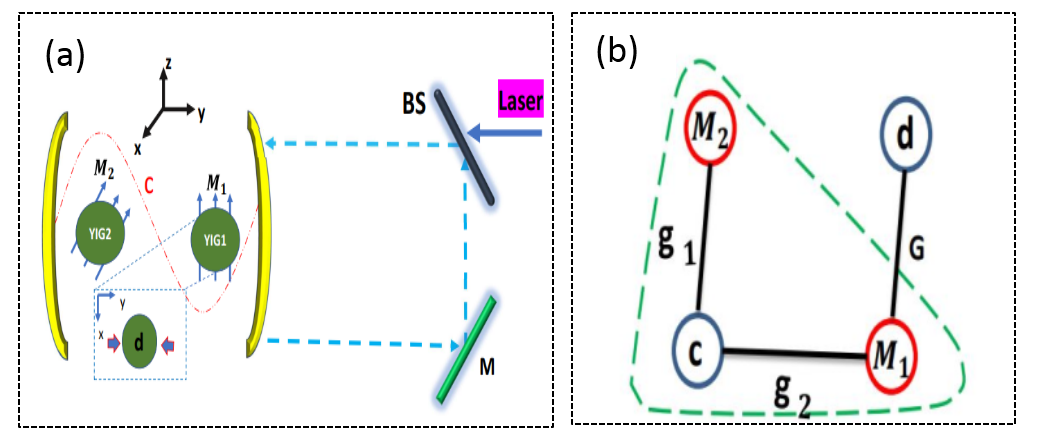}
\caption{(a) This schematic illustrates the hybrid four-mode cavity magnomechanical system. Two YIG samples are placed inside a microwave cavity at points of maximum magnetic field for the cavity mode, while simultaneously being exposed to uniform bias magnetic fields that excite magnon modes within the samples and couple them to the cavity mode. The bias magnetic fields are oriented to activate the magnetostrictive (magnon-phonon) interaction only in one YIG sample (YIG1). This coupling can be further enhanced by directly driving the magnon mode with an external microwave source (not shown). The cavity is also driven by an electromagnetic field through an asymmetric beam splitter (BS) with amplitude $\psi \varepsilon$. We denote the transmission and reflection coefficients by $\psi$ and $\tau$ respectively, and the phase shift generated by the reflectivity of the output field on the mirrors by $\nu$. The photons and magnons of the cavity are coupled by dipole magnetic interaction, and the magnons and phonons are coupled by magnetostrictive interaction.Through an asymmetric beam splitter (BS), a laser light field entering the cavity is split asymmetrically. The output field is completely reflected on the M mirror, and the beam splitter
sends some of the output field into the cavity. (b) Interactions among the subsystems. The cavity mode is linearly coupled to the two magnon modes, \(M_1\) and \(M_2\), with coupling constants $\mathbf{g_1}$ and $\mathbf{g_2}$, respectively. Additionally, magnon mode \(M_1\) is coupled to the mechanical mode \(d\) through the nonlinear magnetostrictive interaction with an effective coupling rate \(\mathbf{G}\). This magnetostrictive interaction leads to magnomechanical entanglement \cite{li2018magnon}, which can be utilized to entangle the two magnon modes. The other indirect couplings are not shown.}
\label{DOC}
\end{figure}

Our model is based on a hybrid four-modes cavity macromechanical system, as seen in Fig. \ref{DOC}, which consists of a mechanical vibration mode, two magnon modes, and a microwave cavity mode. The cavity is associated with a coherent feedback protocol.  The coupling between magnons and cavity photons is mediated by magnetic dipole interactions. The two magnon modes are represented by the collective motion of numerous spins in two macroscopic YIG spheres. This setup, consisting of two YIG spheres without the involvement of a mechanical mode, has been utilized to study magnon dark modes \cite{zhang2015magnon}. The mechanical mode is the YIG crystal's deformation vibration mode caused by the magnetostrictive force. The nature of the magnetostrictive interaction varies depending on the resonance frequencies of the magnon and phonon modes \cite{kittel1958interaction}. Since the frequency of the mechanical mode is much lower than that of the magnon mode, the magnon-phonon coupling is generally weak \cite{zhang2016cavity}. However, this coupling can be significantly enhanced by applying a strong microwave field \cite{li2018magnon, wang2018bistability}. The magnomechanical coupling strength depends on the direction of the bias magnetic field \cite{chen2016mechanical}. We adjust the directions of the two bias magnetic fields  so that the magnetostrictive interaction is effectively activated in only one of the spheres. The system’s Hamiltonian is expressed as $(\hbar = 1)$
\begin{equation}
 \label{eq1}
	\mathcal{H}=\omega_c c^{\dagger}c+\displaystyle\sum_{j=1,2}\left(\omega_{M_j}M^{\dagger}_j M_j\right)+ \frac{\omega_{d}}{2}(\hat{p}^2+\hat{q}^2)
	+\displaystyle\sum_{j=1,2}\mathbf{g_j}(c+c^{\dagger})(M^{\dagger}_j+M_j)+\mathbf{G_0}M_1^{\dagger}M_1\hat{q}\nonumber\\
	+\mathbf{i}\Omega\left(M_1^{\dagger}e^{-i\omega_{0}t}-M_1e^{i\omega_{0}t}\right) \nonumber+ \Lambda\nu(c^{\dagger}e^{i\varphi}+ce^{-i\varphi}) \nonumber, 
\end{equation}
the operators \(c\) and \(c^{\dagger}\) (\(M_j\) and \(M_j^{\dagger}\)) are the annihilation and creation operators for the cavity mode (magnon modes); respectively, they satisfy the commutation relation \([U, U^{\dagger}] = 1\), where \(U = c (M_j)\). The operators \(\hat{q}\) and \(\hat{p}\) (with \([\hat{q}, \hat{p}] = i\)) represent the dimensionless position and momentum quadratures of the mechanical mode. The resonance frequencies of the cavity, magnon, and mechanical modes are denoted by \(\omega_c\), \(\omega_{M_j}\), and \(\omega_d\), respectively. The magnon frequencies, \(\omega_{M_j}\), are determined by the external bias magnetic fields \(\mathbf{H_j}\) through the relation \(\omega_{M_j} = \gamma \mathbf{H_j}\), where \(\gamma / 2\pi = 28\) GHz T\(^{-1}\) represent the gyromagnetic ratio. The coupling rate \(\mathbf{g_j}\) represents the linear coupling between the cavity and the \(j\)th magnon mode, while \(\mathbf{G}_0\) denotes the single-magnon magnomechanical coupling rate. The Rabi frequency \(\Omega = \frac{\sqrt{5}}{4} \gamma \sqrt{N} B_0\) \cite{li2019entangling} quantifies the coupling strength of the drive magnetic field (with amplitude \(B_0\) and frequency \(\omega_0\)) with the first magnon mode. Here, \(N = \rho V\) represents the total number of spins, where \(\rho = 4.22 \times 10^{27} \, \mathrm{m}^{-3}\) is the spin density of YIG, and \(V\) is the volume of the sphere. The first (second) term in the Hamiltonian characterizes the energy of the cavity mode (magnon modes), while the third term represents the energy of the mechanical vibration mode with frequency \(\omega_{d}\). Driving the magnon mode \(M_1\) with a strong microwave field can significantly enhance the effective magnomechanical coupling \cite{li2018magnon}. The driving Hamiltonian is described by the sixth term. Note that this model differs from the one presented in Ref. \cite{li2019entangling}. The last term in equation (\ref{eq1}) describes the optical field generated by the system that is transmitted through the beam splitter \cite{amazioug2020using}, where \(\varphi\) represents the phase of the electromagnetic field. The parameters \(\nu\) and \(\tau\) are the real amplitude transmission coefficients of the beam splitter; these parameters are real, positive, and satisfy the relation \(\nu^2 + \tau^2 = 1\).  
In the frame rotating  at the drive frequency $\omega_{0}$ and by applying the rotating wave approximation (RWA), the expression, ${\mathbf{g_j}}(c+c^{\dagger})(M^{\dagger}_j+M_j)$ becomes $\mathbf{g_j}(cM^{\dagger}_j+c^{\dagger}M_j)$ (valid when $\omega_c, \omega_{M_{j}} \gg \mathbf{g_j}, k_c, k_m$, which is easily satisfied \cite{zhang2016cavity}, the quantum Langevin equations (QLEs) defining the dynamics of the system are provided by
\begin{align}
\dot{c}&=-(i\Delta_{fb}+k_{fb})c-i\mathbf{g_1}M_1-i\mathbf{g_2}M_2- i\nu\Lambda e^{i\varphi}+\sqrt{2k_c}c_{fb}^{in},\\ \nonumber
 \dot{M_1}&=-(i\Delta_{M_1}+k_{M_1})M_1-i\mathbf{G_0}M_1\hat{q}-i\mathbf{g_1}c+\Omega+\sqrt{2k_{M_1}}M_{1}^{in},\\ \nonumber
  \dot{M_2}&=-(i\Delta_{M_2}+k_{M_2})M_2-i\mathbf{g_2}c+\sqrt{2k_{M_2}}M_{2}^{in},\\ \nonumber
\dot{\hat{q}}&=\omega_{d}\hat{p},\\ \nonumber
\dot{\hat{p}}&=-\omega_{d}\hat{q}-\gamma_{d}\hat{p}-\mathbf{G_0}M^{\dagger}_1M_1+\chi,
\end{align}
with  $k_{fb}=k_c(1-2\tau\cos\varphi)$ represents the the effective cavity decay rate, and  $\Delta_{fb} = \Delta_c-2k_c\tau\sin\varphi$ is the effective detuning of the cavity mode. Where $\Delta_c=\omega_c-\omega_{0}$ and  $\Delta_{M_j}=\omega_{M_j}-\omega_{0}$ (j=1,2) are, respectively the detuning of the cavity and magnons modes. The dissipation rates of the cavity, magnons, and mechanical mode are represented by th parameters \(k_c\), \(k_{M_j}\), and \(\gamma_{d}\), respectively. The Rabi frequency $\Omega$ suggests that the magnon mode $M_1$ only receives one driving field application. The operator $C^{in}_{fb}=\tau e^{i\varphi}c^{out}+\nu c^{in}$ is the effective input noise operator in the presence of coherent feedback. Here $c^{in}$ represents the noise operator associated with microwave mode with only non-zero correlations \cite{amazioug2023feedback}. Additionally, the standard input-output relation $c^{out}=\sqrt{2k_c}c-\nu c^{in}$ establishes a relationship between the output field $c^{out}$ and the cavity field $c$, meaning that $C^{in}_{fb}=\tau\sqrt{2k_c}e^{i\varphi}c+c^{in}_{fb}$.  Furthermore, for the cavities, the non-zero coherent feedback correlations qualities of the input noise operators are provided by, whose non-zero correlation functions are (with $\varphi=0$)
 \begin{equation}
 \begin{aligned} 
\langle c^{in}_{fb}(t)c^{in\dagger}_{fb}(t')\rangle &=\{\nu(1-\tau)\}^2[n_c(\omega_c)+1]\delta(t-t'), \\ 
\langle c^{in}_{fb}(t)c^{in\dagger}_{fb}(t')\rangle &=\{\nu(1-\tau)\}^2[n_c(\omega_c)]\delta(t-t').
\end{aligned}
\end{equation}
The magnon mode, which is associated with the noise operator  $M_j^{in}$ ($j=1,2$), has a zero mean and is characterized by the following correlation functions:
$\langle M_j^{in}(t)M_j^{in\dagger}(t')\rangle  =[N_{M_j}(\omega_{M_j})+1]\delta(t-t')$, and $\langle M_j^{in\dagger}(t)M_j^{in}(t')\rangle  =N_{M_j}(\omega_{M_j})\delta(t-t'). $
 The Langevin force
operator $\chi$ is accounting for the mechanical Brownian motion, which is autocorrelated as $\langle\chi(t)\chi(t')+\chi(t')\chi(t)\rangle\approx\gamma_{d}[2N_{d}(\omega_{d})+1]\delta(t-t')$, where we consider a high quality factor $Q_d=\omega_d/\gamma_d\gg 1$ for the mechanical oscillators
to validate the Markov approximation \cite{vitali2007optomechanical}. Here, $N_k(\omega_k)=[\exp(\frac{\hbar\omega_k}{k_bT})-1]^{-1} (k=c,M_j,d), (j=1,2)$ are the thermal photon, magnon, and phonon numbers at equilibrium, respectively.  Here $T$ is the environmental temperature and  $k_b$ being the Boltzmann constant.\\
\section{Linearization and covariant matrix}
\label{sec3}
Since the magnon mode $M_1$ is strongly driven by an external microwave field, and due to the beam splitter interactions between the cavity and the two magnon modes, both the cavity and magnon modes have large amplitudes, $|\langle M_j\rangle|,|\langle c\rangle|\gg 1~~(j=1,2)$. This allows us to linearize the system's dynamics around the steady-state values by expressing each operator as the sum of the steady-state average and a fluctuation quantum operator, i.e., we decompose each operator $\psi$ as follows: $\psi=\langle \psi\rangle+\delta \psi \quad(\psi=c,M_j,\hat{q},\hat{p}),(j=1,2)$, and by neglecting small second-order fluctuation terms. Since we are particularly interested in the quantum correlation properties of the two magnon modes, we focus on the dynamics of the system's quantum fluctuations. The linearized quantum Langevin equations (QLEs) that describe the fluctuations of the system quadratures $ [\delta X,\delta Y,\delta x_1,\delta y_1,\delta x_2,\delta y_2,\delta \hat{q},\delta \hat{p}]$, with  $\delta X=\frac{(\delta c+\delta c^{\dagger})}{\sqrt{2}},\delta Y=\frac{i(\delta c^{\dagger}-\delta c)}{\sqrt{2}},\delta x_j=\frac{(\delta M_j+\delta M^{\dagger}_j)}{\sqrt{2}},\delta y_j=\frac{i(\delta M^{\dagger}_j-\delta M_j)}{\sqrt{2}}$, can be written in the following compact matrix  

\begin{equation}
\dot{\mu}(t)=\Gamma \mu(t)+\lambda (t),
\end{equation}
where $\mu(t)=[\delta X(t),\delta Y(t),\delta x_1(t),\delta y_1(t),\delta x_2(t),\delta y_2(t),\delta \hat{q}(t),\delta \hat{P}(t)]^{T}$ is the vector of the quadrature fluctuations, and $\lambda(t)=[\sqrt{2k_{c}}X^{in}(t),\sqrt{2k_{c}}Y^{in}(t),\sqrt{2k_{M_1}}x_1^{in}(t),\sqrt{2k_{M_2}}y_1^{in}(t),\sqrt{2k_{M_2}}x_2^{in}(t),\sqrt{2k_{M_2}}y_2^{in}(t),0,\chi(t)]^T$ is the vector of the input noises entering the system. The drift matrix $\Gamma$ is given by
\[
\Gamma=
\begin{pmatrix}
-k_{fb}&\Delta_{fb}&0&\mathbf{g_1}&0&\mathbf{g_2}&0&0\\
-\Delta_{fb}&-k_{fb}&\mathbf{-g_1}&0&\mathbf{-g_2}&0&0&0\\
0&\mathbf{g_1}&-k_{M_1}&\tilde{\Delta}_{M_1}&0&0&-\mathbf{G}&0\\
\mathbf{-g_1}&0&-\tilde{\Delta}_{M_1}&-k_{M_1}&0&0&0&0\\
0&\mathbf{g_2}&0&0&-k_{M_2}&\Delta_{M_2}&0&0\\
\mathbf{-g_2}&0&0&0&-\Delta_{M_2}&-k_{M_2}&0&0\\
0&0&0&0&0&0&0&\omega_{d}\\
0&0&0&\mathbf{G}&0&0&-\omega_{d}&-\gamma_{d}
\end{pmatrix},
\]
where $\tilde{\Delta}_{M_1}={\Delta}_{M_1}+ G_0 \langle q \rangle $, with $\langle \hat{q}\rangle =-\mathbf{G_0}|\langle M_1\rangle|^2/\omega_{d}$. The effective magnomechanical coupling rate, which is typically complex, is represented by  $\mathbf{G}=i\sqrt{2}\mathbf{G_{0}}\langle M_1\rangle$, here $
	\left\langle M_1\right\rangle \simeq-\left(\mathbf{g_1} \langle c \rangle+i \Omega\right) / \tilde{\Delta}_{M_{1}}$. 
The average $\langle \hat{P}\rangle$, $\langle c\rangle$, and $\langle M_2\rangle$     are given by  
\begin{equation}
	\begin{aligned}
		\langle \hat{p}\rangle &= 0, \quad \langle c\rangle \simeq  \frac{ ( \nu\Lambda e^{i\varphi} \tilde{\Delta}_{M_1}+ i \mathbf{g_1} \Omega) \Delta_{M_2} }{ \Delta_{f b} \tilde{\Delta}_{M_{1}}\Delta_{M_2}-\mathbf{g_1}^{2}\Delta_{M_2}+ \mathbf{g_2}^{2} \tilde{\Delta}_{M_1}} \quad \text{and}  \quad		 
		\left\langle M_2\right\rangle \simeq-\left(\mathbf{g_2} \langle c \rangle \right) / {\Delta}_{M_{2}}.
	\end{aligned}
\end{equation}
The above expressions of $\langle c\rangle$, $\langle M_j\rangle$ ($j=1,2$) are achieved under the condition that $|\Delta_{fb}|,|\tilde{\Delta}_{M_1}|,|\Delta_{M_2}|\gg k_{fb},k_{M_1},k_{M_2}$, \cite{li2019entangling}, and $g_1\Omega\gg \nu\Lambda\tilde{\Delta}_{M_{1}}$. In this instance, $\langle c\rangle$ and $\langle M_j\rangle$ ($j=1,2$) are pure imaginary numbers. According to the Routh-Hurwitz criterion \cite{dejesus1987routh}, the system is stable. The drift matrix $\Gamma$ is given under this condition. In fact, we will show later that. 
The stationary CM $V$ can be easily determined by solving the Lyapunov equation \cite{vitali2007optomechanical}
\begin{equation}
	\label{5}
\Gamma V+V\Gamma^{T}=-\mathcal{F}
\end{equation}
where $\mathcal{F}$ is the diffusion matrix, it is described as $\mathcal{F}_{ij}=\langle \lambda_i(t)\lambda_j(t')+\lambda_i(t')\lambda_j(t)\rangle/[2\delta(t-t')]$ and given by $\mathcal{F}=diag[k_c(2N_c+1)\nu^2(1-\tau)^2,k_c(2N_c+1)\nu^2(1-\tau)^2,k_{M_1}(2N_{M_1}+1),k_{M_1}(2N_{M_1}+1),k_{M_2}(2N_{M_2}+1),k_{M_2}(2N_{M_2}+1),0,\gamma_{d}(2N_{d}+1)]$. Once the CM for the steady-state system is obtained from the previously mentioned calculation, we can examine the bipartite and tripartite entanglement properties among the different four-mode. By removing the rows and columns linked to the uninteresting modes from $V$, the interested modes state can be retrieved and their entanglement properties examined once the CM of the system fluctuations has been obtained. In the following, the Gaussian bipartite/tripartite entanglement is quantified using the logarithmic negativity and the minimum residual contangle, respectively.

\section{Bipartite and tripartite quantum correlations}
\label{sec4}
\subsection{Logarithm negativity}
The covariance matrix $\sigma_{PQ}$ of two magnon can be derived from equation (\ref{5}) 
\begin{equation}
	\label{segm}
\sigma_{PQ}=
\begin{pmatrix}
P & W \\
W^{T} & Q
\end{pmatrix}
\end{equation}
the autocorrelations of the two modes are represented by the $2 \times 2$ sub-matrices $P$ and $Q$ in Eq. (\ref{5}), and their cross-correlations are defined by the $2 \times 2$ sub-matrix $W$ in  Eq. (\ref{5}). We define the logarithmic negativity $L_N$ as a proof of the entanglement in bipartite subsystem in CV system. It is defined as \cite{adesso2007entanglement}
\begin{equation}
\label{ln}
L_{N}=\max[0,-log(2\Sigma)]
\end{equation}
with $\Sigma=\frac{\sqrt{\chi-(\chi^2-4\det\sigma_{PQ})^{1/2}}}{\sqrt{2}}$ being the smallest symplectic eigenvalue of partial transposed covariance matrix of two magnon, with $\chi=\det P+\det Q-2\det W$. The logarithmic negativity is related by $\Sigma$, then if ($\Sigma<1/2$) thus the two subsystems are entangled. 
\subsection{Quantum steering}
We define the Gaussian quantum steerability as asymmetric property between two entangled subsystems (two magnon modes), the steerability of Bob(P) by Alice(Q) ($P\rightarrow Q$) or ($Q\rightarrow p$) can be measured as \cite{hamza}
\begin{equation}
S^{P \rightarrow Q}(\sigma_{PQ})=max\left[0,\frac{1}{2}ln\left(\frac{\det P}{4\det\sigma_{PQ}}\right)\right],
\end{equation}
the steerability of Alice by Bob [$S^{P \rightarrow Q}(\sigma_{PQ})$] can be obtained by swapping the roles of $P$ and $Q$. Can be procured by  exchanging the roles of $P$ and $Q$. Interestingly, the opposite is not necessarily true: a steerable state is always a non-separable state. As a result, we have two options between $P$ and $Q$: firstly if $S^{Q\rightarrow P}=S^{P\rightarrow Q}=0$  Alice can’t steer Bob and vice versa even if they are entangled
(i.e., no-way steering), secondly is $S^{P\rightarrow Q}>0$ and  $S^{Q\rightarrow P}=S^{P\rightarrow Q}=0$ or $S^{Q\rightarrow P}>0$ as one-way steering, i.e., Alice can steer Bob but Bob can’t steer Alice and vice versa, and thirdly if $S^{P\rightarrow Q}=S^{Q\rightarrow P}>0$ Alice can steer Bob and vice versa (i.e., two-way steering). In addition, the measurement of Gaussian Steering is always bounded by the entanglement. In addition to examining the two mode Gaussian state’s asymmetric steerability, we also present steering asymmetry, given by \cite{hamza}
\begin{equation}
S(PQ)=|S^{P\rightarrow Q}-S^{Q\rightarrow P}|.
\end{equation}
\subsection{Gaussian Geometric Discord}
A measure of all non-classical correlations in Gaussian states of continuous variable systems is the Gaussian geometric discord ($\mathcal{D}_G$). It measures correlations in two-mode Gaussian states by extending the geometric measure of quantum discord. For the two-mode squeezed thermal states, the $\mathcal{D}_G$ can be explicitly expressed as follows \cite{cheng2021tripartite}
\begin{equation}
\mathcal{D}_G=\frac{1}{4(pq-w^2)}-\frac{9}{\left(2\sqrt{4 pq-  3 w^2}+2\sqrt{pq}\right)^2},
\end{equation}
with $p$, $q$ and $w$ are the elements of covariance matrix $\sigma_{PQ}$ in Eq.( \ref{segm}), i.e., $P=\diag (p,p)$, $Q= \diag (q,q)$ and $W=\diag (w,-w)$.
 
\subsection{Genuine tripartite entanglement}
We employ the residual contangle to examine the system's tripartite entanglement $\mathcal{R}$ \cite{adesso2007entanglement} as quantitative measure. Hence, for discrete-variable tripartite entanglement, contangle is the CV equivalent of tangle \cite{amazioug2023enhancement}. The
quantification of tripartite entanglement is given by the minimum residual contangle \cite{amazioug2023enhancement,adesso2006continuous}
\begin{equation}
\mathbf{R}_{min}\equiv min[\mathbf{R}^{c|M_1d},\mathbf{R}^{M_1|cd},\mathbf{R}^{d|cM_1}],
\end{equation}
where $\mathbf{R}^{ijk} $ is the residual contangle, with $\mathbf{C}_{v|w}$ is  the contangle of subsystems of $v$ and $w$ (w contains one or two modes), it is the squared logarithmic negativity and a proper entanglement monotone $\mathbf{C}_{v|w}\equiv E_{v|w}^{2}$. Besides, a nonzero minimum residual contangle $\mathbf{R}_{min}$, exhibit the existence of
genuine tripartite entanglement in the system. $\mathbf{R}^{i|jk}$ is similar to the Coffman-Kundu-Wootters monogamy
inequality \cite{amazioug2023enhancement} hold for the system of three qubits. $\mathbf{R}^{ijk}$ is provided by 
\begin{equation}
 \mathbf{R}^{ijk}=C_{i|jk}-C_{i|j}-C_{i|k}\geq 0  \quad   (i,j,k=c,M_1,d),
\end{equation}
when $w$ represents a single mode  $E_{v|w}$ ( $ E_{c|d}$, $E_{c|M1}$, and  $E_{d|M1} $), we apply Eq. (\ref{ln}). When the   $w$ represents two modes $E_{v|w}$(  $ E_{c|dM1}$, $E_{d|cM1}$, and   $E_{M1|cd} $), in this case the $\Sigma$ in Eq. (\ref{ln}) is defined by 
\begin{equation}
\Sigma =\min \operatorname{eig}\left[i \hat{\Omega}_3 \tilde{\mathcal{V}}\right]
\end{equation}
where 
$\hat{\Omega}_3$ and $\tilde{\mathcal{V}}$ are defined, respectively, as
$
\hat{\Omega}_3=\oplus_{j=1}^3 i \sigma_y, \quad \sigma_y=\left(\begin{array}{cc}
	0 & -i \\
	i & 0
\end{array}\right)
$ and $ \tilde{\mathcal{V}}=\mathcal{R}_{i \mid j k} V \mathcal{R}_{i \mid j k}, \quad(i, j, k=c, d, M_1)
$ where $\quad \mathcal{R}_{c \mid d M1}=\sigma_z\oplus 1 \oplus 1, \quad \mathcal{R}_{d \mid c M1}=$ $1\oplus \sigma_z \oplus 1$, and $\mathcal{R}_{M1 \mid c d}=1\oplus 1 \oplus \sigma_z$, with $\sigma_z=\diag(1,-1)$.
\section{Results and Discusion}
\label{sec6}

In this section, we present the results and delve into the evolution of quantum correlations within the system and show how the presence of feedback enhances entanglement. We have employed experimentally attainable parameters \cite{zhang2016cavity, li2019entangling}: $\omega_c/2\pi=10$ GHz, $\omega_b/2\pi=10$ GHz, $\gamma_d/2\pi=10^2$ Hz, $k_c/2\pi=10$ MHz, $k_{M_1(M_2)}=k_c$, $\mathbf{g_1}/2\pi=3.2$ MHz, $\mathbf{g_2}/2\pi=2.6$ MHz, $\mathbf{G}/2\pi=4.8$ MHz, the temperature $T=10$ mK and $\varphi=0$. 
\begin{figure}[ht!]
\centering
\begin{tabular}{ll}
\includegraphics[width=0.5\linewidth]{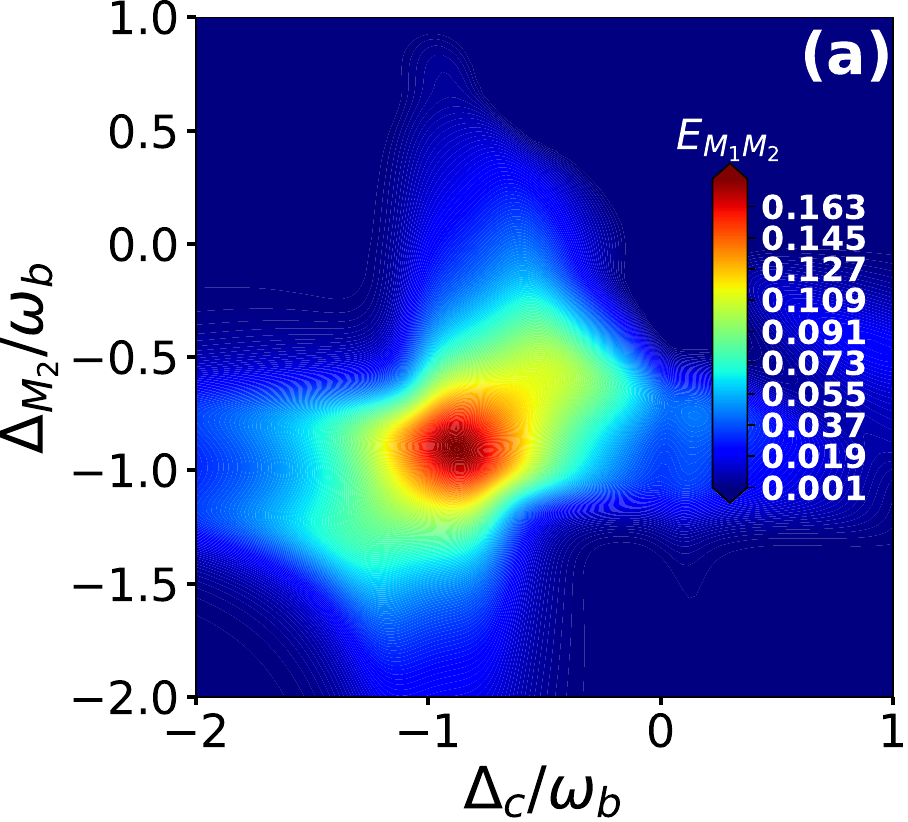}
\includegraphics[width=0.5\linewidth]{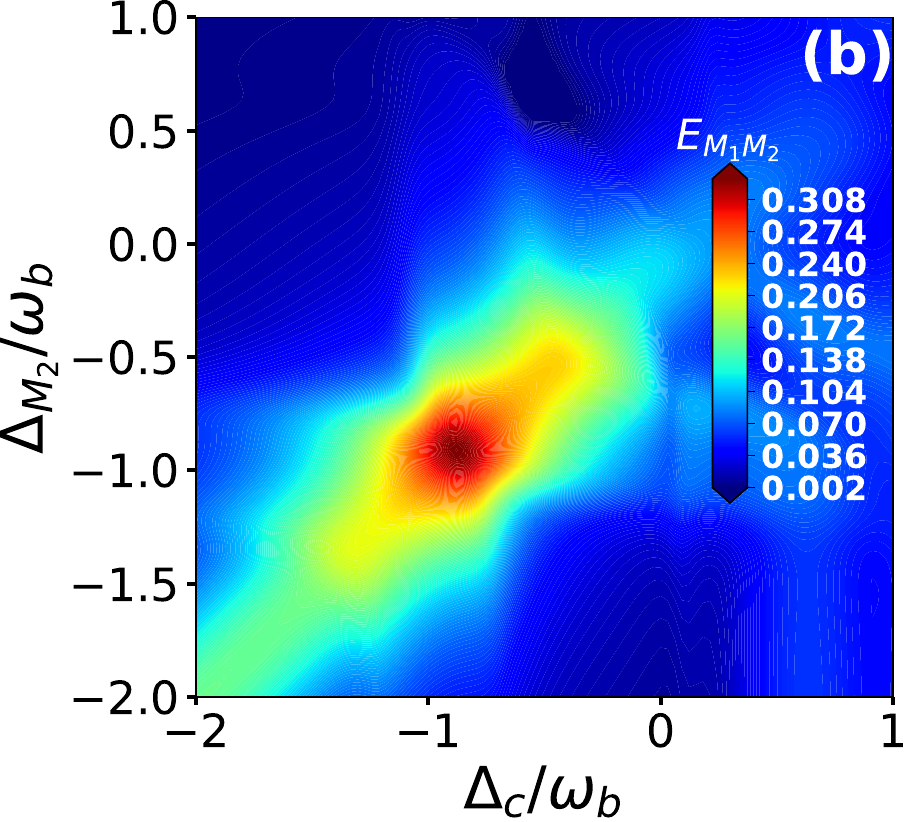}\\ 
\hspace{0.66cm}\includegraphics[width=0.48\linewidth]{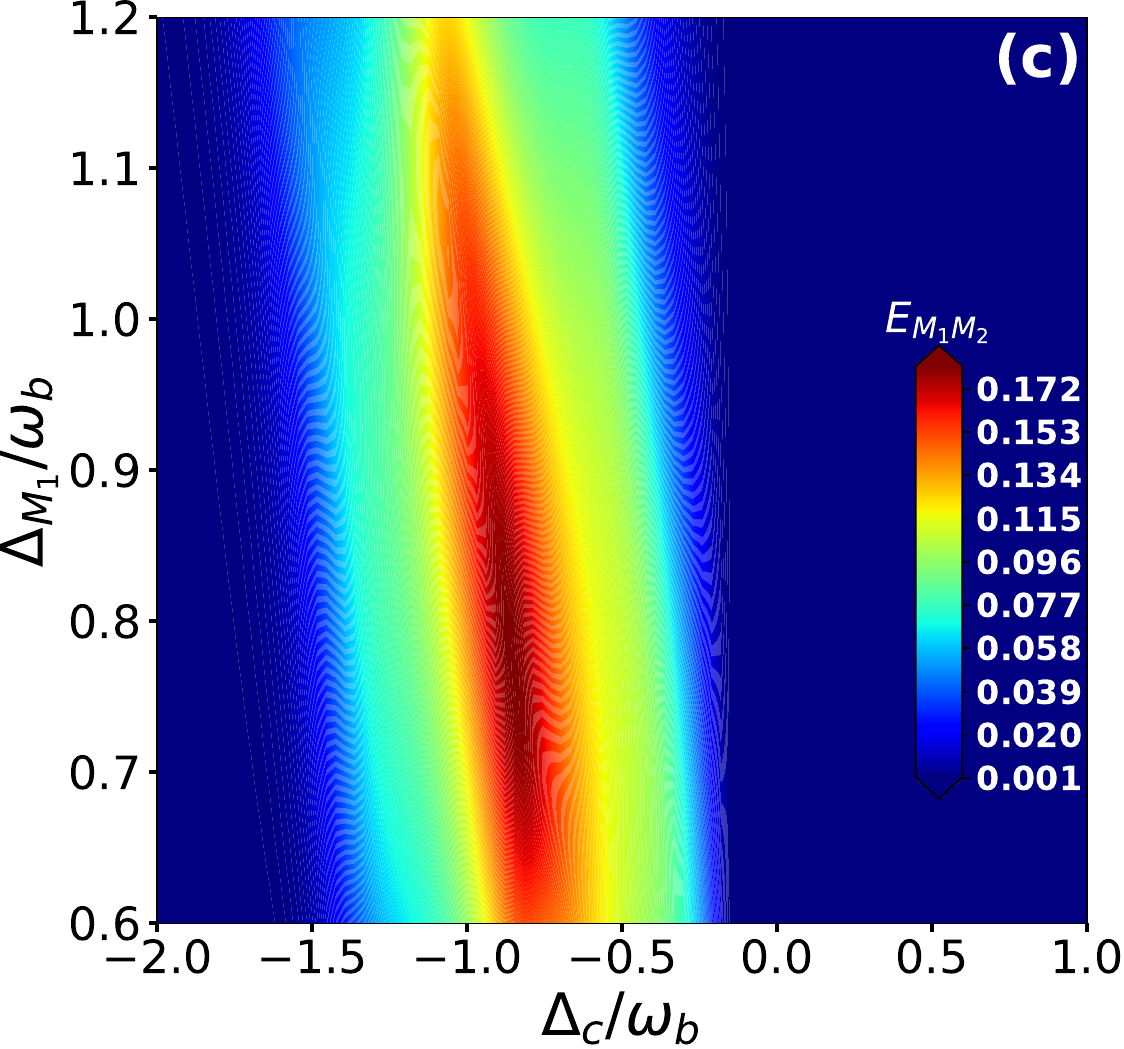} 
\hspace{0.5cm}\includegraphics[width=0.47\linewidth]{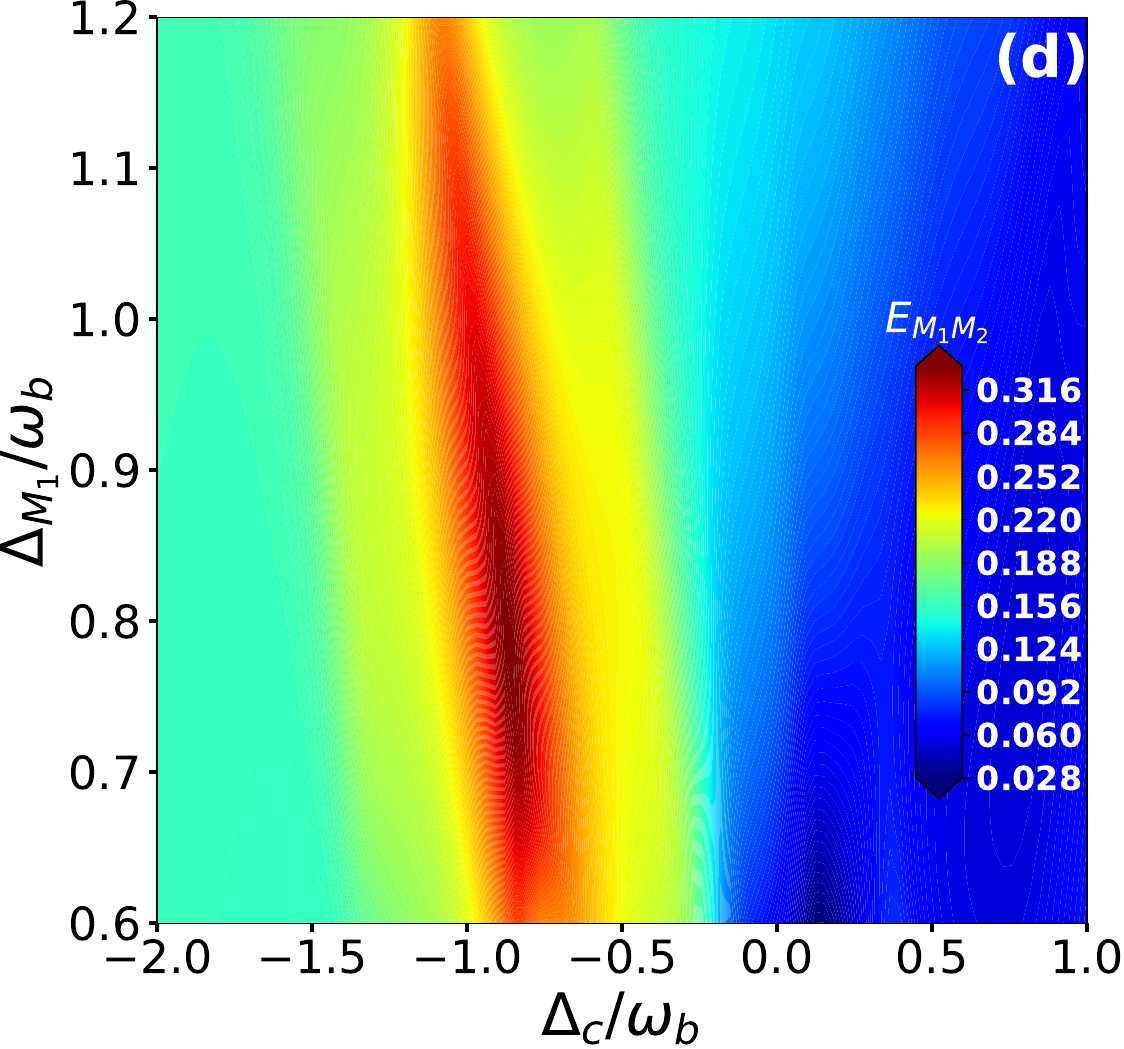}
\end{tabular}
\caption{Density plot of bipartite entanglement $E_{M_1M_2}$ between two magnon modes as a function of (a-b) $\Delta_c$ and $\Delta_{M_2}$, (c-d) $\Delta_c$ and $\Delta_{M_1}$. We take $\Delta_{M_1}=0.85\omega_b$ in (a-b) and $\Delta_{M_2}=\Delta_c$ in (c-d). With in Figs. (a-c) $\tau=0$ and in Figs. (b-d) $\tau=0.3$.}
\label{fig:gf}
\end{figure}

First, in Fig. \ref{fig:gf}(a-b), we explore the variation of bipartite entanglement $E_{M_{1} M_{2}}$ in steady state as a function of normalized detunings $\Delta_{M_{2}}/\omega_d$ and $\Delta_c/\omega_b$. The entanglement reaches its maximum when $\Delta_{M_2}\approx\Delta_c\approx-\omega_b$, indicating resonance between the magnon mode and the cavity. This is also the case for generating squeezed states of magnons by driving the cavity with a squeezed microwave field \cite{li2019entangling}. In this region, $E_{M_1M_2}$ attains its maximum value of $E_{M_1M_2}=0.31$ when $\tau=0.3$ (see Fig. \ref{fig:gf}(b)), and $E_{M_1M_2}=0.16$ when $\tau=0$ \cite{li2019squeezed}. We observe that coherent feedback boost the entanglement of the two magnons, by comparing the results shown in Fig. \ref{fig:gf}(b) with respect to Fig. \ref{fig:gf}(a). \\
Next, the Fig. \ref{fig:gf}(c-d) exhibits the evolution of bipartite entanglement $ E_{M_1M_2}$ with respect to $\Delta_{M_1}/\omega_b$ and $\Delta_c/\omega_b$. We observe a significant improvement of the magnon-magnon entanglement through the coherent feedback loop, as depicted in Fig. \ref{fig:gf}(d). This improvement can be attributed to the re-injection of photons into the cavity. Furthermore, the entanglement reaches its optimal value, $E_N=0.32$ for $\Delta_c=-\omega_b$ and $\tau=0.3$. Moreover the optimel value of the entanglement is $E_N=0.172$ for $\Delta_c=-\omega_b$ and $\tau=0$ \cite{li2019squeezed} (see Fig. \ref{fig:gf}(c)).

\begin{figure}[htbp]
\centering
\begin{tabular}{ll}
\includegraphics[width=0.48\linewidth]{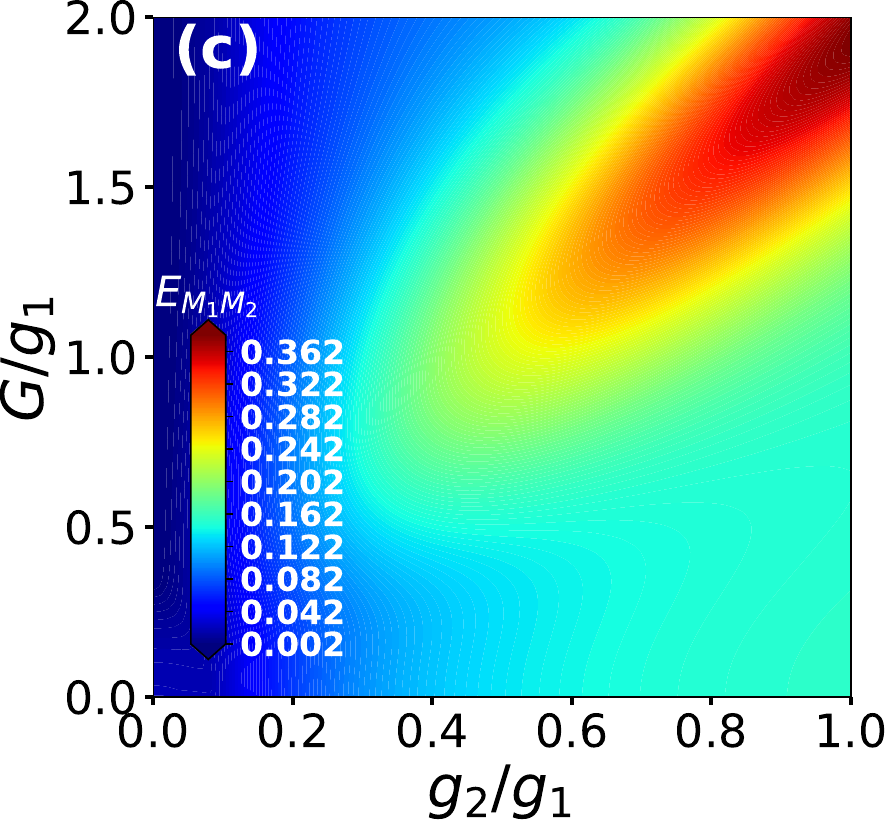} 
\includegraphics[width=0.52\linewidth]{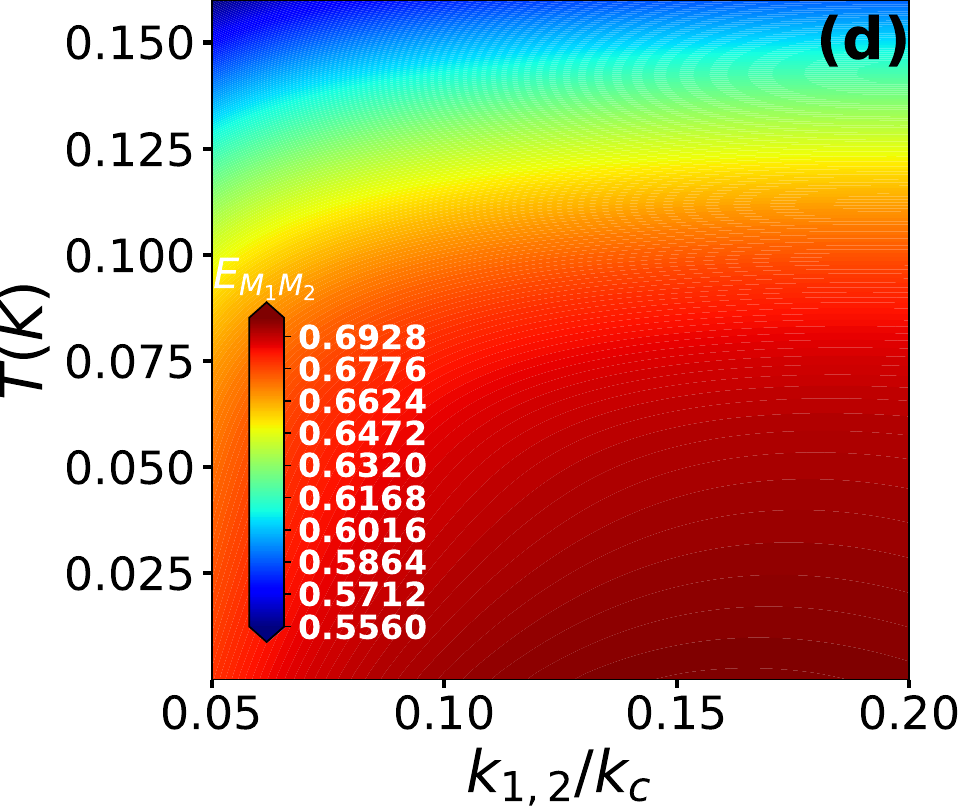} 
\end{tabular}
\caption{(c) Density plot of bipartite entanglement $E_{M_1M_2}$ between two magnon modes as function of the ratios of $g_2/g_1$ and $G/g_1$ ($g_1$ fixed) $\tau=0.3$. (d) Density plot of bipartite entanglement $E_{M_1M_2}$ between two magnon modes as function of the temperature $T$ the ratio $k_{1,(2)}/k_c$ with $\tau=0.3$.}
\label{fig:gf2}
\end{figure}

We show in Fig. \ref{fig:gf2}(c), the density plot of bipartite entanglement between two magnon modes as function of  the ratio  $G/g_1$ and $g_2/g_1$. The plot reveals that entanglement reaches its maximum when $G=2g_1$ and $g_1=g_2$ considering a fixed $g_1$. Moreover, coherent feedback enhances entanglement compared to the results highlighted in \cite{li2019squeezed}. Additionally, logarithmic negativity $E_{M_1M_2}$ is increased by both $G$ and $g_2$. Notably, when $G=0$ or $g_2=0$, logarithmic negativity $E_{M_1M_2}$ becomes zero. This implies that the indirect coupling between the two magnons, mediated by the magnon-phonon interaction, is essential for their entanglement.

In Fig. \ref{fig:gf}(d), we show the variation of logarithmic negativity $E_{M_1M_2}$ as a function of the temperature $T$ and the ratio of the two magnon modes and cavity dissipations $k_{1,2}/k_c$. We observe that as temperature increases, magnon-magnon entanglement rapidly degrades due to decoherence effects induced by thermal noise. Conversely, the entanglement proves to be quite robust ($E_{M_1M_2}>0.69$) against variations in the two magnon rates, persisting up to approximately $T\ll$ 50 mK for $k_{1,2}=2$ MHz. The augmentation of magnon dissipation contributes to lowering the temperature threshold for enhancing magnon-magnon entanglement in the presence of a coherent feedback loop.

\begin{figure}[!h]

  (a)\includegraphics[width=0.45\linewidth]{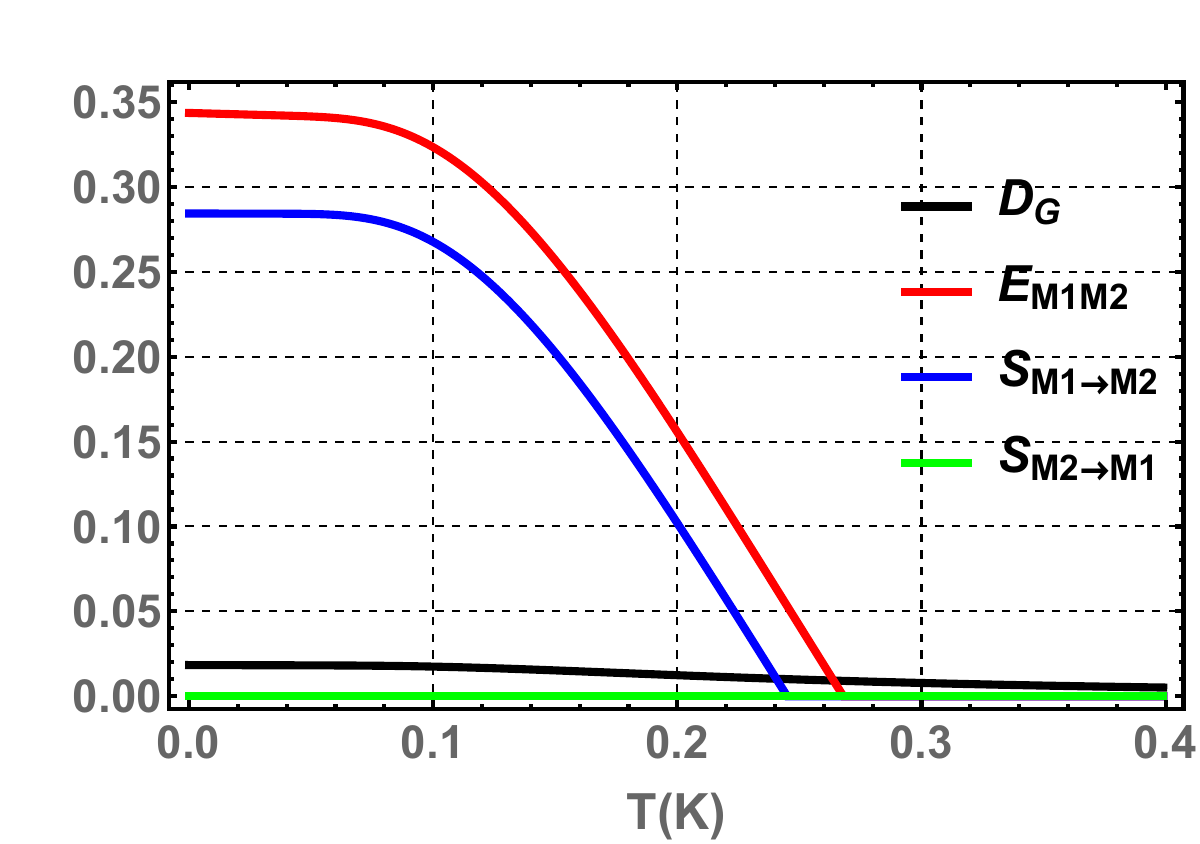} 
  (b)\includegraphics[width=0.45\linewidth]{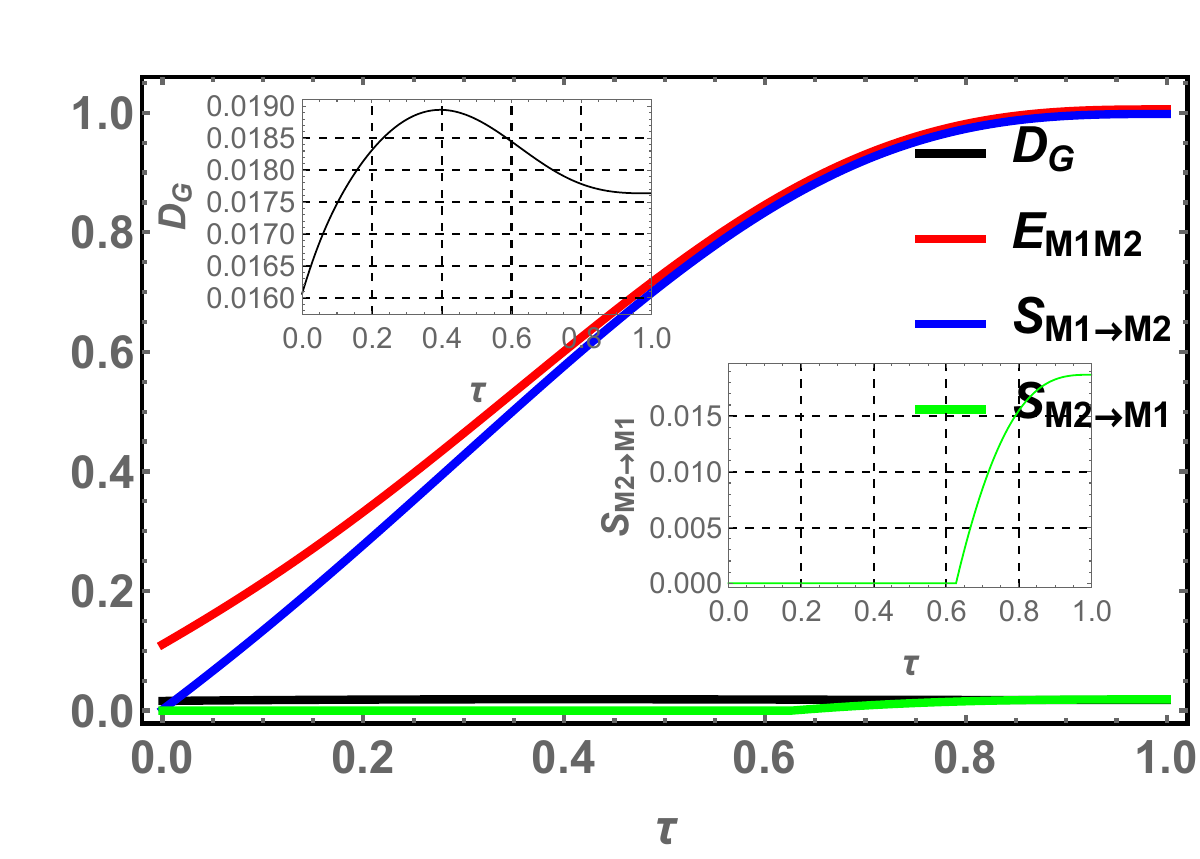} 
\begin{center}
\caption{Plot of $\mathcal{D}_G$  and  the  bipartite the entanglement of two magnons modes $E_{M_1M_2}$, the Gaussian quantum steering $S_{M_1\rightarrow M_2}$, and $S_{M_2\rightarrow M_1}$, as a function of temperature $T$ ($\tau$ = 0.2) (a), and reflectivity parameter $\tau$ (b). We take $\mathbf{G}/2\pi=4.8$ MHz, $\Delta_c=-\omega_b$, $\Delta_b=0.9\omega_b$, $k_{fb}=5k_{M_{1(2)}}$ and $k_{fb}/2\pi= 3$ MHz.} 
\label{fig:Emm}
\end{center}
\end{figure}

In Fig. \ref{fig:Emm}, we show the behavior of Gaussian geometric discord $\mathcal{D}_{G}$, bipartite entanglement $E_{M_1M_2}$, and the steerabilities $S_{M_1 \rightarrow M_2}$ and $S_{M_2 \rightarrow M_2}$ between the two magnon modes, as functions of the temperature $T$ (Fig. \ref{fig:Emm}(a)) and reflectivity $\tau$ (Fig. \ref{fig:Emm}(b)) in the presence of coherent feedback. We observe that all three functions decrease monotonically with increasing temperature, a phenomenon attributable to decoherence \cite{zurek2003decoherence}. The logarithmic negativity drops to zero around $T_{\text{max}} \approx 250$ mK, while $\mathcal{D}_{G}$ remains nonzero even at 300 mK, demonstrating non-classical correlations beyond entanglement. At lower temperatures ($T\to 0$), steerability is maximum ($S_{M_1\rightarrow M_2}\approx 0.27$) and varies steadily with increasing temperature. However, once the temperature exceeds 70 mK, steerability decreases monotonically and vanishes at temperatures above 220 mK. Notably, $S_{M_2 \rightarrow M_2}$ remains zero across all temperature intervals, indicating one-way steerability from $M_1$ to $M_2$. This asymmetry in quantum steerability is explained by the fact that magnon 1 can steer magnon 2 only when the shared state is entangled, while magnon 2 cannot steer magnon 1. This panel demonstrates that increasing temperature can destroy both entanglement and steerability, but $\mathcal{D}_{G}$ exhibits robust behavior against temperature.

The Fig. \ref{fig:Emm}(b) shows the evolution of the entanglement ($E_{M_1M_2}$), steerability ($S_{M_1\rightarrow M_2}$, $S_{M_2\rightarrow M_1}$) and $\mathcal{D}_G$ between two magnon modes as a function the reflectivity parameter $\tau$. We observe that $\mathcal{D}_G$, $S_{M_1\rightarrow M_2}$, $S_{M_2\rightarrow M_1}$, and $E_{M_1M_2}$ increase monotonically with increasing $\tau$. $\mathcal{D}_G$ reaches a maximum of approximately 0.019 at $\tau=0.4=$ and then decreases as reflectivity exceeds 0.4. Both entanglement $E_{M_1M_2}$ and steerability $S_{M_1\rightarrow M_2}$ increase for lower values of $\tau$, reaching their maximum values at $\tau=1$. Additionally, steerability $S_{M_1\rightarrow M_2}$ emerges when $\tau\geq 0.6$ and continues to increase with increasing $\tau$. This indicates the emergence of two-way steering for $\tau\geq 0.6$ (as shown in the inset figure of Fig. \ref{fig:Emm}(b)).

From the above results, we conclude that the reflectivity parameter $\tau$ significantly affects all three forms of quantum correlation. Gaussian geometric discord is more robust against thermal noise compared to entanglement and steerability. Additionally, we demonstrate the asymmetric nature of the steering system. Steerability is a stronger form of quantum correlation than entanglement but weaker than Bell nonlocality. While all steerable states are entangled, not all entangled states are steerable. Steerability remains bounded by entanglement.

\begin{figure}
\includegraphics[width=0.45\linewidth]{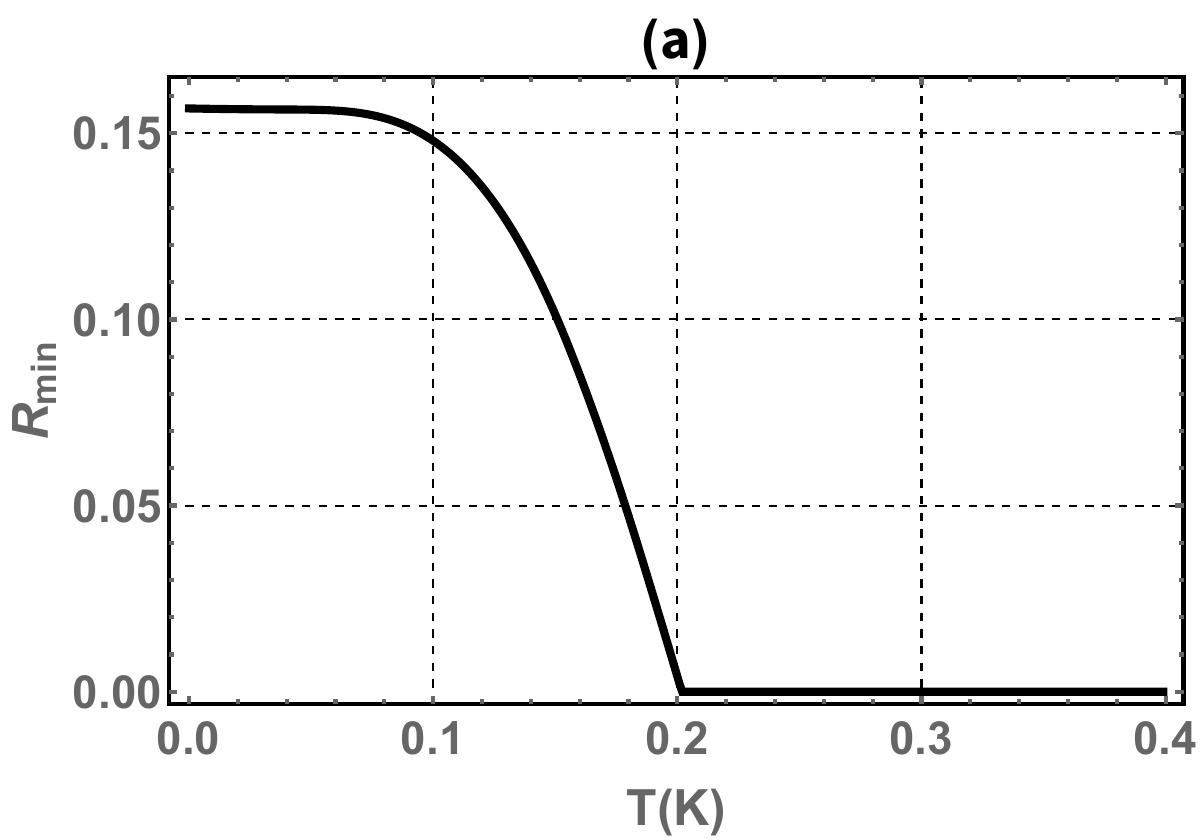} 
\includegraphics[width=0.45\linewidth]{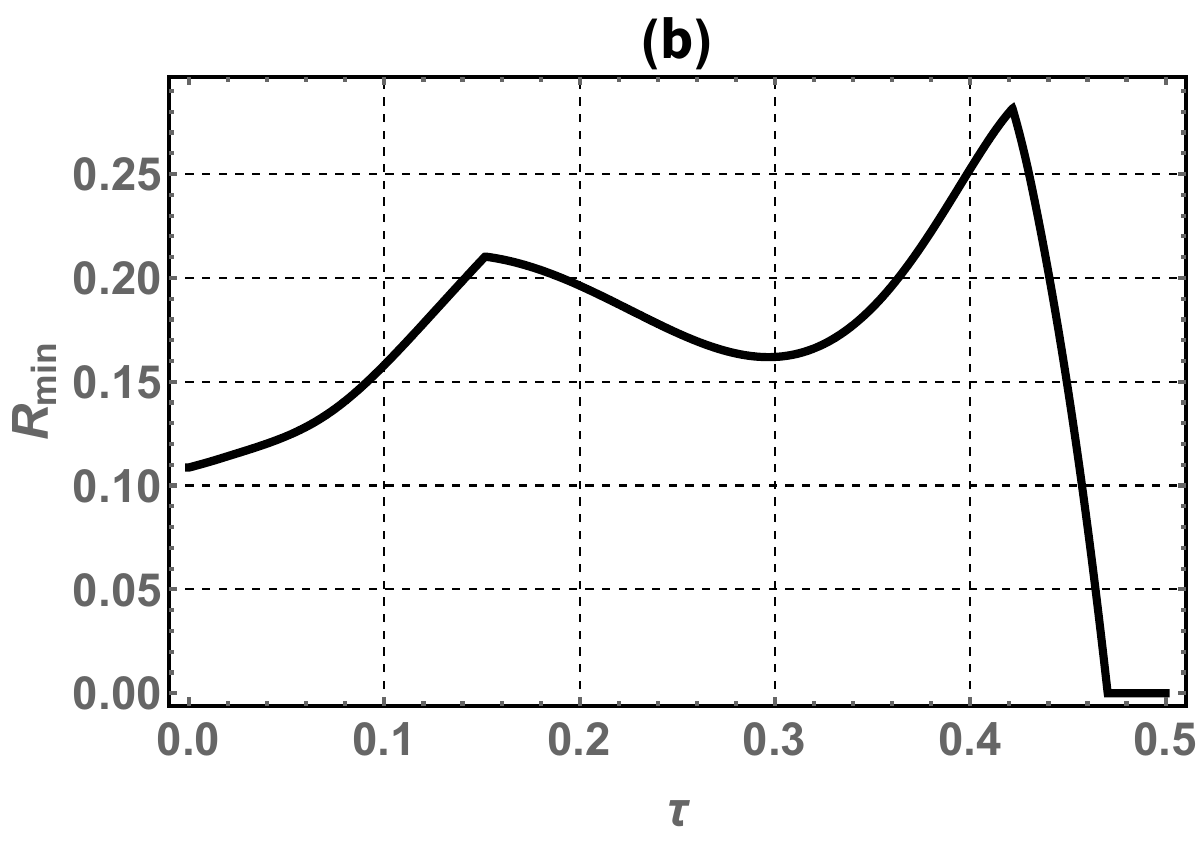} 
\begin{center}
\caption{Plot of tripartite entanglement in terms of the minimum residual contangle $\mathcal{R}_{min}$ as function of (a) temperature $T$ with $\tau=0.1$ and (b) reflectivity $\tau$. The parameters are as in Fig. \ref{fig:Emm}.} 
\label{fig:residual}
\end{center}
\end{figure}

 Then, the Fig. \ref{fig:residual}(a) shows the evolution of tripartite entanglement as measured by the minimum residual contangle $R_{min}$ agains the temperature $T$ for $\tau=0.1$. This panel depicts the residual tripartite entanglement among the three modes (photon-magnon1-phonon). We observe that tripartite entanglement increases as temperature $T$ decreases. When the temperature exceeds 80 mK, the contangle start to diminishes rapidly and vanishes at 200 mK due to decoherence effects. Higher temperature contribute to destroy tripartite entanglement.

Finally, the  Fig. \ref{fig:residual}(b) presents the minimum residual contangle $R_{min}$ as a function of reflectivity $\tau$. At low values of $\tau$, we observe an enhancement of $R_{min}$, indicating that the coherent feedback loop boosts tripartite entanglement. However, the tripartite entanglement degrades rapidly after reaching its maximum value at $\tau=0.41$, eventually vanishing around $\tau=0.48$.

\section{Conclusion}
\label{sec7}

In summary, we have investigated the enhancement of bipartite (magnon1-magnon2) and tripartite (photon-magnon1-phonon) entanglement in a cavity magnomechanical system through coherent feedback. We employed logarithmic negativity to quantify in steady-state the entanglement between the two magnons. Gaussian quantum steering was used to measure the steerability, revealing its dependence on entanglement. Gaussian geometric discord was employed to quantify all quantum correlations between the two magnons. Our findings demonstrate that coherent feedback can significantly enhances Gaussian quantum steering, entanglement, and Gaussian geometric discord. We observed one-way steering between magnon1 and magnon2 for $k_{fb}=k_{{1(2)}}$, while one-way and two-way steering emerged for $k_{fb}=5k_{{1(2)}}$ through adjustments to the reflectivity parameter. The genuine tripartite entangled state was quantified using minimum residual contangle. Both bipartite and tripartite quantum correlations were found to be fragile under thermal noise, but coherent feedback effectively mitigated these effects. The feasibility of our scheme is discussed, employing recent technologies.

\end{document}